\documentclass[12pt]{article}
\usepackage{amsmath,amssymb,amscd,amsthm,latexsym}

\font\block=msbm10
\def\C{\hbox{\block\char'0103}}
\def\Z{\hbox{\block\char'0132}}

\font\gotic=eufm10
\def\g{\hbox{\gotic\char'0147}}

\font\new=eusm10
\def\A{\hbox{\new\char'0101}}
\def\P{\hbox{\new\char'0120}}

\def\W{\hbox{\new\char'0127}}

\font\yes=msam10
\def\Q{\hbox{\yes\char'0003}}

\begin{document}

\noindent
{\Large Embedding of the Lie superalgebra $D(2, 1 ; \alpha)$
into the Lie superalgebra of pseudodifferential symbols on $S^{1|2}$}

\vskip 0.3in
{\large Elena Poletaeva}
\vskip 0.1in

{\it School of Mathematics, Institute for Advanced Study,
Princeton, NJ 08540 and}

{\it Department of Mathematics,
University of Texas-Pan American,}

{\it Edinburg, TX 78539 (permanent address)}

{\it Electronic mail:} elena@math.ias.edu and elenap$@$utpa.edu

\vskip 0.4in

{\footnotesize \noindent {\bf Abstract.}
We obtain an embedding of a one-parameter family of exceptional simple Lie superalgebras
$D(2, 1 ; \alpha)$ into the
Lie superalgebra of pseudodifferential symbols on the supercircle $S^{1|2}$.
Correspondingly, there is an embedding of $D(2, 1 ; \alpha)$ into
a nontrivial central extension of the derived contact superconformal algebra  $K'(4)$ realized in terms of $4\times 4$
matrices over a Weyl algebra.}

\vskip 0.4in
\noindent
{\bf I. Introduction}
\vskip 0.2in

Recall that $D(2, 1 ; \alpha)$ with $\alpha \in \C\backslash
\lbrace 0, -1 \rbrace$ is a one-parameter family of classical simple Lie superalgebras of dimension 17 (Ref. 1).
The bosonic part of $D(2, 1 ; \alpha)$ is
$sl(2)\oplus sl(2)\oplus sl(2)$, and the action of
$D(2, 1 ; \alpha)_{\bar 0}$ on $D(2, 1 ; \alpha)_{\bar 1}$
is the  product of $2$-dimensional representations.
These superalgebras are also denoted in the literature by
$\Gamma (\sigma_1, \sigma_2,  \sigma_3)$, where $\sigma_i$ are nonzero complex numbers
such that $\sigma_1+\sigma_2+\sigma_3 = 0$ (Refs. 2, 3).
$\Gamma (\sigma_1, \sigma_2, \sigma_3)\cong D(2, 1 ; \alpha)$, where
$\alpha = \sigma_1/{\sigma_2}$.
In Ref. 4 M. G{\"u}naydin gave a differential operator realization of the action of
$D(2, 1 ; \alpha)$ on a family of superspaces with 2 bosonic and 2 fermionic coordinates.

In this work we consider
$\Gamma (2, -1 - \alpha,  \alpha - 1)$ as a one-parameter family of
deformations of the Lie superalgebra $spo(2|4)\cong osp(4|2)$ embedded into the
Poisson superalgebra $P(4)$ of pseudodifferential symbols on the supercircle $S^{1|2}$
with even variable $t$ and odd variables $\xi_1$ and $\xi_2$.
$P(4) = P\otimes \Lambda(4)$, where $P$ is
the Poisson algebra of functions on the cylinder ${T}^*S^1\backslash S^1$
(which are formal Laurent series in $\tau = {\partial\over {\partial t}}$
along the fibres, with coefficients periodic in $t$),
and $\Lambda(4) = \Lambda(\xi_1, \xi_2, \eta_1, \eta_2)$ is the Grassmann algebra,
see Ref. 5.
If $\alpha = 0$, then $\Gamma (2, -1,  - 1)\cong spo(2|4)$,
and it is naturally embedded into $P(4)$.
Note that $spo(2|4)_{\bar 1}$ is spanned by the zero modes
of the fermionic fields of two copies of a Lie superalgebra, which is isomorphic to the derived superalgebra $S'(2, 0)$ of
divergence-free derivations
of $\C[t, t^{-1}]\otimes\Lambda (\xi_1, \xi_2)$. We  obtain one copy, if we
identify ${\partial\over {{\partial{\xi_i}}}}$ with $\eta_i$
 for $i = 1, 2$.  To obtain the other copy
we interchange
$\xi_i$ with $\eta_i$ in all formulas.
Then using  the Schwimmer-Seiberg's deformation $S'(2, \alpha)$ (see Refs. 6 and 7)
of each copy of  this superalgebra, we
embed $\Gamma (2, -1 - \alpha, \alpha - 1)$ into
$P(4)$ for each $\alpha \in \C$. There is also an
embedding of $\Gamma (2, -1 - \alpha, \alpha - 1)$ into
the family of Lie superalgebras of pseudodifferential symbols
$P_h(4)$, where $h\in (0, 1]$, which contracts to $P(4)$.

Note that $S'(2, \alpha)$ is spanned by 4 bosonic
and 4 fermionic fields, and it is a subsuperalgebra of the derived contact superconformal algebra $K'(4)$, which is spanned by 8 bosonic
and 8 fermionic fields (Refs. 7, 8, 9 and 10).
$K'(4)$ is also known to physicists as the (centerless) ``big  $N = 4$ superconformal algebra'' (Refs. 11 and 12).
We have shown in Ref. 5 that there exists an embedding of one of three independent nontrivial central extensions
$\hat{K}'(4)$ of $K'(4)$ into $P_h(4)$ for each $h\in (0, 1]$.
Note that this central extension is different from the one that corresponds to the Virasoro cocycle.
Associated to these embeddings, there are spinor-like irreducible representations of
$\hat{K}'(4)$ in the superspaces
$V^{\mu} = t^{\mu}\C[t, t^{-1}]\otimes \Lambda(\xi_1, \xi_2)$,
where $({\partial\over {\partial t}})^{-1}$ acts as an antiderivative.
This requires that $\mu\in\C\backslash\Z$.
Nevertheless,  a representation of $\hat{K}'(4)$
in $V^{\mu}$ is well-defined even if $\mu = 0$. In this case we obtain a realization
of $\hat{K}'(4)$ in terms of $4\times 4$ matrices over
a Weyl algebra $\W = \sum_{i\geq 0}\A d^i$, where $\A = \C[t, t^{-1}]$ and $d = t{\partial\over {\partial t}}$.
Then we describe
$\Gamma (2, -1 -\alpha,  \alpha - 1)$ as a subsuperalgebra of $\hat{K}'(4)$
for each $\alpha \in \C$.

In Ref. 14 (see also Ref. 13) we used the similar approach to realize the
exceptional $N = 6$ superconformal algebra, which is spanned by 32 fields
(Refs. 9, 10 and 15-18),
as a subsuperalgebra of $8\times 8$ matrices over a Weyl algebra.
This realization is  analogous to the realization, given by
C. Martinez and E. I. Zelmanov in Refs. 19 and 20, where they used a different method.

Note that the affine superalgebra
$\hat{D}(2, 1 ; \alpha)$ is closely related to the big  $N = 4$ superconformal algebra (see Ref. 21).
It is an interesting problem to realize $\hat{D}(2, 1 ; \alpha)$
in terms of pseudodifferential symbols and matrices
over a Weyl algebra. We would also like to find such realizations for
the exceptional Lie superalgebra $F(4)$ (Ref. 1).

\vskip 0.2in
\noindent
 {\bf II. Superconformal algebras}
\vskip 0.2in

A {\it superconformal algebra} is a complex
Lie superalgebra $\g$ such that
\hfil\break
1) $\g$ is simple,
\hfil\break
2) $\g$ contains the Witt  algebra $Witt = der \C [t, t^{-1}] =
 \oplus_{n\in \Z}\C L_n$
 with the well-known commutation relations
$$[L_n, L_m] = (m - n)L_{n+m} \eqno (2.1)$$
 as a subalgebra,
\hfil\break
3) $ad L_0$ is diagonalizable with finite-dimensional eigenspaces:
$$\g = \oplus_i\g_i,\quad \g_i =
\lbrace x\in\g \mid [L_0, x] = ix \rbrace, \eqno (2.2)$$
so that $dim$$\g_i < C$, where $C$ is a constant independent of $i$,
see Refs. 7, 8, 22 and 23.

Let $\Lambda(2N)$ be the Grassmann algebra in $2N$ variables
$\xi_1, \ldots, \xi_N, \eta_1, \ldots, \eta_N $, and
let $\Lambda(1, 2N) =\C [t, t^{-1}]\otimes \Lambda (2N)$ be an associative
superalgebra with natural multiplication and with the following parity
of generators: $p(t) = \bar{0}$, $p(\xi_i) = p(\eta_i) = \bar{1}$
for $i = 1, \ldots, N$. Let
$W(2N)$ be the Lie superalgebra of all superderivations of
$\Lambda(1, 2N)$.
Let $\partial_t$, $\partial_{\xi_i}$ and $\partial_{\eta_i}$
stand for $\partial\over {\partial t}$,
$\partial\over {\partial \xi_i}$ and
$\partial\over {\partial \eta_i}$, respectively.
Every $D\in W(2N)$ is represented by a differential operator,
$$D = f\partial_t + \sum_{i=1}^N (f_i \partial_{\xi_i} + g_i \partial_{\eta_i}), \eqno (2.3)$$
where $f, f_i, g_i \in \Lambda (1, 2N)$.

The Lie superalgebra $W(2N)$ contains
 a one-parameter family of Lie
superalgebras $S(2N, \alpha)$.
By definition
$$S(2N, \alpha) = \lbrace D \in W(2N) \mid
\hbox{Div}(t^{\alpha}D) = 0\rbrace \hbox { for } \alpha \in \C,\eqno (2.4)$$
see Refs. 6 and 7.
Recall that
$$\hbox{Div}(D) =
\partial_t(f) + \sum_{i=1}^N ((-1)^{p(f_i)}\partial_{\xi_i} (f_i) + (-1)^{p(g_i)}\partial_{\eta_i} (g_i)). \eqno (2.5)$$
Let $S'(2N, \alpha) =  [S(2N, \alpha), S(2N, \alpha)]$ be the derived
superalgebra.
Assume that $N \geq 1$. If ${\alpha}\not\in\Z$, then
$S(2N, \alpha)$ is simple, and if
${\alpha} \in \Z$, then
$S'(2N, \alpha) $ is a  simple ideal of $S(2N, \alpha)$ of codimension one
defined from the exact sequence,
$$0\rightarrow S'(2N, \alpha)\rightarrow
 S(2N, \alpha)\rightarrow \C t^{-\alpha} {\xi_1} \cdots  {\eta_N}
\partial_t\rightarrow 0.\eqno (2.6)$$
Notice that
$$S(2N, \alpha)\cong S(2N, \alpha + n) \hbox{ for } n\in\Z. \eqno (2.7)$$
There exists, up to equivalence, one nontrivial 2-cocycle on $S'(2N, \alpha)$
if and only if $N = 1$, see Ref. 7.
The corresponding central extension
$\hat{S}'(2, 0)$ is also called the  ``$N = 4$ superconformal algebra'' (Refs. 11, 12,  9 and 22).
Let
$$\lbrace L_n, E_n, H_n, F_n, h_n, p_n, x_n, y_n\rbrace_{n\in \Z} \eqno (2.8)$$
be the following basis of $S'(2, 0)$:
\begin{equation*}
\begin{aligned}
&L_n = -t^n(t\partial_t + {1\over 2}(n+1)
(\xi_1\partial_{\xi_1} + \eta_1\partial_{\eta_1}),\\
&E_n = t^n\eta_1\partial_{\xi_1},\quad
H_n = t^n(\eta_1\partial_{\eta_1} - \xi_1\partial_{\xi_1}),\quad
F_n = t^n\xi_1\partial_{\eta_1},\\
&h_n = t^n\eta_1\partial_t - nt^{n-1}\xi_1\eta_1\partial_{\xi_1},\quad
p_n = t^{n+1}\partial_{\eta_1},\\
&x_n = t^{n+1}\partial_{\xi_1},\quad
y_n = t^n\xi_1\partial_t + nt^{n-1}\xi_1\eta_1\partial_{\eta_1}.\\
\end{aligned}
\tag{2.9}
\end{equation*}
The 2-cocycle in $\hat{S}'(2, 0)$ is given as follows:
\begin{equation*}
\begin{aligned}
&c(L_n, L_k) = {1\over {12}}(n^3 - n)\delta_{n+k, 0},\\
&c(E_n, F_k) = {1\over {6}}n\delta_{n+k, 0},\quad
c(H_n, H_k) = {1\over {3}}n\delta_{n+k, 0},\\
&c(h_n, p_k) = -{1\over {6}}(n^2 - n)\delta_{n+k, 0},\quad
c(x_n, y_k) = -{1\over {6}}(n^2 + n)\delta_{n+k, 0}.\\
\end{aligned}
\tag{2.10}
\end{equation*}

\vskip 0.2in
\noindent
{\bf III. Poisson superalgebra}
\vskip 0.2in

 The {\it Poisson algebra $P$ of pseudodifferential symbols on
the circle} is formed by the formal series
$$A(t, \tau) = \sum_{-\infty}^na_i(t)\tau^i,\eqno (3.1)$$
where $a_i(t)\in \C[t, t^{-1}]$, and the even variable $\tau$ corresponds to $\partial_t$, Refs. 24-27.
The Poisson bracket is defined as follows:
$$\lbrace A(t, \tau), B(t, \tau)\rbrace = \partial_{\tau}A(t, \tau)\partial_tB(t, \tau) - \partial_tA(t, \tau)\partial_{\tau}B(t, \tau).\eqno (3.2)$$
An associative algebra
$P_h$, where $h\in  (0, 1]$, is a deformation of $P$. The multiplication in $P_h$ is given as follows:
$$A(t, \tau)\circ_hB(t, \tau) = \sum_{n\geq 0 }{h^n\over {n!}}
\partial_{\tau}^nA(t, \tau)\partial_t^nB(t, \tau).\eqno (3.3)$$
The Lie algebra structure on the vector space $P_h$ is given by
$$[A, B]_h = A\circ_hB - B\circ_hA,\eqno (3.4)$$
so that
$$\hbox{lim}_{h\rightarrow 0}{1\over h}[A, B]_h = \lbrace A, B\rbrace.\eqno (3.5)$$
The {\it Poisson superalgebra of pseudodifferential symbols on $S^{1|N}$} is
$P(2N) = P\otimes \Lambda(2N)$. The Poisson bracket is defined as follows:
$$\lbrace A, B\rbrace = \partial_{\tau}A\partial_tB - \partial_tA\partial_{\tau}B +
(-1)^{p(A)+1}\sum_{i = 1}^N(\partial_{\xi_i}A\partial_{\eta_i}B + \partial_{\eta_i}A\partial_{\xi_i}B).\eqno (3.6)$$
Let $\Lambda_h(2N)$ be an associative superalgebra with generators
$\xi_1, \ldots, \xi_N, \eta_1, \ldots, \eta_N$ and relations
$$\xi_i\xi_j = - \xi_j\xi_i,\quad \eta_i\eta_j = -\eta_j\eta_i, \quad \eta_i\xi_j = h\delta_{i, j} - \xi_j\eta_i.\eqno (3.7)$$
Let $P_h(2N) = P_h\otimes\Lambda_h(2N)$ be a superalgebra with the product given by
$$(A_1\otimes X)(B_1\otimes Y) = (A_1\circ_hB_1)\otimes (XY),\eqno (3.8)$$
 where
$A_1, B_1 \in P_h$ and $X, Y \in \Lambda_h(2N)$.
The Lie bracket of $A = A_1\otimes X$ and $B = B_1\otimes Y$  is
$$[A, B]_h = AB - (-1)^{p(A)p(B)}BA,\eqno (3.9)$$
and (3.5) holds.
$P_h(2N)$ is called the {\it Lie superalgebra of pseudodifferential symbols on $S^{1|N}$},
see Refs. 5 and 13.

\vskip 0.2in
\noindent
{\bf IV. Superalgebras $\Gamma(\sigma_1, \sigma_2, \sigma_3)$}
\vskip 0.2in

Recall the definition of $\Gamma(\sigma_1, \sigma_2, \sigma_3)$, see Refs. 1, 2, 3.
Let $\g = \g_{\bar{0}} \oplus \g_{\bar{1}}$ be a Lie superalgebra, where
$\g_{\bar{0}} = sp(\psi_1)\oplus sp(\psi_2)\oplus sp(\psi_3)$ and
$\g_{\bar{1}} = V_1\otimes V_2\otimes V_3$, where
$V_i$ are $2$-dimensional vector spaces, and
$\psi_i$ is a non-degenerate skew-symmetric form on $V_i$, $i = 1, 2, 3$.
A representation of $\g_{\bar{0}}$ on $\g_{\bar{1}}$ is the tensor product
of the standard representations of $sp(\psi_i)$ in $V_i$.
Consider $sp(\psi_i)$ - invariant bilinear mapping
$$\P_i: V_i\times V_i \rightarrow sp(\psi_i), \quad i = 1, 2, 3,\eqno (4.1)$$
given by
$$\P_i(x_i, y_i)z_i = \psi_i(y_i, z_i)x_i - \psi_i(z_i, x_i)y_i\eqno (4.2)$$
for all $x_i, y_i, z_i\in V_i$.
Let $\P$ be a mapping
$$\P:\g_{\bar{1}}\times \g_{\bar{1}}\rightarrow \g_{\bar{0}}\eqno (4.3)$$
given by
\begin{equation*}
\begin{aligned}
&\P(x_1\otimes x_2\otimes x_3, y_1\otimes y_2\otimes y_3) =\\
&\sigma_1\psi_2(x_2, y_2)\psi_3(x_3, y_3)\P_1(x_1, y_1) + \\
&\sigma_2\psi_1(x_1, y_1)\psi_3(x_3, y_3)\P_2(x_2, y_2) + \\
&\sigma_3\psi_1(x_1, y_1)\psi_2(x_2, y_2)\P_3(x_3, y_3)
\end{aligned}
\tag{4.4}
\end{equation*}
for all $x_i, y_i \in V_i, i = 1, 2, 3$,
where $\sigma_1, \sigma_2, \sigma_3$ are some complex numbers.
The Jacobi identity is satisfied if and only if
$\sigma_1 + \sigma_2 + \sigma_3 = 0$. In this case $\g$ is denoted by
$\Gamma(\sigma_1, \sigma_2, \sigma_3)$.
Superalgebras $\Gamma(\sigma_1, \sigma_2, \sigma_3)$ and
$\Gamma(\sigma_1', \sigma_2', \sigma_3')$ are isomorphic if and only if there exists a nonzero element $k\in \C$ and a permutation $\pi$ of the set $\lbrace 1, 2, 3\rbrace$ such that
$$ \sigma_i' = k\cdot\sigma_{\pi i} \hbox{ for } i = 1, 2, 3.$$
Superalgebras $\Gamma(\sigma_1, \sigma_2, \sigma_3)$ are simple if
and only if $\sigma_1, \sigma_2, \sigma_3$ are all different from zero (see Ref. 3).
Note that $\Gamma(\sigma_1, \sigma_2, \sigma_3)\cong D(2, 1 ; \alpha)$, where
$\alpha = \sigma_1/{\sigma_2}$.

\vskip 0.1 in

{\bf Theorem 4.1:}
Let $\Gamma_{\alpha}$, where $\alpha \in \C$,  be the Lie superalgebra spanned by
the following elements in $P(4)$:
\begin{equation*}
\begin{aligned}
&E_{\alpha}^1 =  t^2, \quad
F_{\alpha}^1 =  \tau^2 - 2\alpha t^{-2}\xi_1\xi_2\eta_1\eta_2, \quad
H_{\alpha}^1 =  t\tau,\\
&E_{\alpha}^2 = \xi_1\xi_2,\quad
F_{\alpha}^2 = \eta_1\eta_2, \quad
H_{\alpha}^2 =  \xi_1\eta_1 + \xi_2\eta_2, \\
&E_{\alpha}^3 =  \xi_1\eta_2, \quad
F_{\alpha}^3 = \xi_2\eta_1,\quad
H_{\alpha}^3 =  \xi_1\eta_1 - \xi_2\eta_2,\\
&T_{\alpha}^1 =  t\eta_1, \quad
T_{\alpha}^2 =  t\eta_2 ,\quad
T_{\alpha}^3 =  t\xi_1, \quad
T_{\alpha}^4 =  t\xi_2,\\
&D_{\alpha}^1 =  \tau\xi_1 + \alpha t^{-1}\xi_1\xi_2\eta_2,\quad
D_{\alpha}^2 = \tau\xi_2 - \alpha t^{-1}\xi_1\xi_2\eta_1,\\
&D_{\alpha}^3 =  \tau\eta_1 + \alpha t^{-1}\xi_2\eta_1\eta_2,\quad
D_{\alpha}^4 =  \tau\eta_2 - \alpha t^{-1}\xi_1\eta_1\eta_2.
\end{aligned}
\tag{4.5}
\end{equation*}
Then $\Gamma_{\alpha}\cong \Gamma(2, -1 - \alpha,\alpha - 1)$ for each $\alpha \in \C$.

\vskip 0.1 in
\noindent
{\it Proof.}
Consider the following fields:
\begin{equation*}
\begin{aligned}
&H_{n, \alpha}^1 =  t^{n+1}\tau, \quad
H_{n, \alpha}^2 =  t^n(\xi_1\eta_1 + \xi_2\eta_2), \\
&E_{n, \alpha}^3 =  t^n\xi_1\eta_2, \quad
F_{n, \alpha}^3 = t^n\xi_2\eta_1,\quad
H_{n, \alpha}^3 =  t^n(\xi_1\eta_1 - \xi_2\eta_2),\\
&T_{n, \alpha}^1 =  t^{n+1}\eta_1, \quad
T_{n, \alpha}^2 =  t^{n+1}\eta_2 ,\quad
T_{n, \alpha}^3 =  t^{n+1}\xi_1, \quad
T_{n, \alpha}^4 =  t^{n+1}\xi_2,\\
&D_{n, \alpha}^1 =  t^n\tau\xi_1 + (\alpha + n) t^{n-1}\xi_1\xi_2\eta_2,\quad
D_{n, \alpha}^2 = t^n\tau\xi_2 - (\alpha +n) t^{n-1}\xi_1\xi_2\eta_1,\\
&D_{n, \alpha}^3 =  t^n\tau\eta_1 + (\alpha +n) t^{n-1}\xi_2\eta_1\eta_2,\quad
D_{n, \alpha}^4 =  t^n\tau\eta_2 - (\alpha +n) t^{n-1}\xi_1\eta_1\eta_2.
\end{aligned}
\tag{4.6}
\end{equation*}
Set
$$L_{n, \alpha}^1 =
H_{n, \alpha}^1  + {1\over 2}(\alpha + n + 1)H_{n, \alpha}^2.\eqno(4.7)$$
Let
$$S_{\alpha}^1\subset P(4)\eqno(4.8)$$
be defined as follows:
$$S_{\alpha}^1 = \hbox{Span}(L_{n, \alpha}^1, E_{n, \alpha}^3, F_{n, \alpha}^3,
H_{n, \alpha}^3,
T_{n, \alpha}^1, T_{n, \alpha}^2, D_{n, \alpha}^1,  D_{n, \alpha}^2).\eqno(4.9)$$
Then
$$S_{\alpha}^1\cong S'(2, \alpha) \hbox{ for  all } \alpha\in \C \eqno (4.10) $$
Note that (4.8) is the restriction of the obvious embedding of $W(2)$ into $P(4)$.
We obtain an embedding of the second copy of $S'(2, \alpha)$:
$$S_{\alpha}^2\subset P(4)\eqno (4.11)$$
by interchanging $\xi_i$ with $\eta_i$ in all formulas.
Thus
$$S_{\alpha}^2 = \hbox{Span}(L_{n, \alpha}^2, E_{n, \alpha}^3, F_{n, \alpha}^3,
H_{n, \alpha}^3,
T_{n, \alpha}^3, T_{n, \alpha}^4, D_{n, \alpha}^3,  D_{n, \alpha}^4),\eqno (4.12)$$
where
$$L_{n, \alpha}^2 = H_{n, \alpha}^1  - {1\over 2}(\alpha + n + 1)H_{n, \alpha}^2.\eqno (4.13)$$
Let $spo(2|4)\cong osp(4|2)$ be a Lie superalgebra which preserves an
even nondegenerate superskew-symmetric form on the $(2|4)$-dimensional superspace.
Note that if $\alpha = 0$, then the zero modes of the fermionic fields
$$T_{n, \alpha}^i \hbox{ and } D_{n, \alpha}^i \hbox{ for } i = 1, 2, 3, 4\eqno (4.14)$$
span $spo(2|4)_{\bar 1}$, hence these elements generate
$\Gamma_0\cong spo(2|4)\cong \Gamma (2, -1, -1)$.
Analogously, for each $\alpha \in \C$, the zero modes of the fields (4.14)
generate $\Gamma_{\alpha}$ and it  is isomorphic to $\Gamma (2, -1 - \alpha, \alpha -1)$.

Explicitly an  isomorphism $\varphi: \Gamma(2, -1 -\alpha, \alpha - 1) \rightarrow  \Gamma_{\alpha}$ is given as follows.
Let
\begin{equation*}
\begin{aligned}
&V_1 =\hbox{Span}(e_1, e_2) , \quad
V_2   = \hbox{Span}(f_1, f_2), \quad
V_3 = \hbox{Span}(h_1, h_2),
\end{aligned}
\end{equation*}
and
\begin{equation*}
\begin{aligned}
&\psi_1(e_1, e_2) = - \psi_1 (e_2, e_1) = 1,\\
& \psi_2(f_1, f_2) = - \psi_2(f_2, f_1) = 1,\\
&\psi_3(h_1, h_2) = - \psi_3(h_2, h_1) = 1.
\end{aligned}
\tag{4.15}
\end{equation*}
Then
\begin{equation*}
\begin{aligned}
&\varphi(\P_1(e_1, e_1)) = -E_{\alpha}^1, \quad
\varphi(\P_1(e_2, e_2)) = -F_{\alpha}^1,\quad
\varphi(\P_1(e_1, e_2)) = -H_{\alpha}^1,\\
&\varphi(\P_2(f_1, f_1)) = - 2F_{\alpha}^2, \quad
\varphi(\P_2(f_2, f_2)) = - 2E_{\alpha}^2,\quad
\varphi(\P_2(f_1, f_2)) =  H_{\alpha}^2,\\
&\varphi(\P_3(h_1, h_1)) = - 2F_{\alpha}^3, \quad
\varphi(\P_3(h_2, h_2)) =  2E_{\alpha}^3,\quad
\varphi(\P_3(h_1, h_2)) =  H_{\alpha}^3,\\
&\varphi(e_1\otimes f_1\otimes h_1) = {\sqrt 2}iT_{\alpha}^1, \quad
\varphi(e_1\otimes f_1\otimes h_2) = {\sqrt 2}iT_{\alpha}^2, \\
&\varphi(e_1\otimes f_2\otimes h_1) = -{\sqrt 2}iT_{\alpha}^4, \quad
\varphi(e_1\otimes f_2\otimes h_2) = {\sqrt 2}iT_{\alpha}^3,\\
&\varphi(e_2\otimes f_1\otimes h_1) = {\sqrt 2}iD_{\alpha}^3, \quad
\varphi(e_2\otimes f_1\otimes h_2) = {\sqrt 2}iD_{\alpha}^4, \\
&\varphi(e_2\otimes f_2\otimes h_1) = -{\sqrt 2}iD_{\alpha}^2, \quad
\varphi(e_2\otimes f_2\otimes h_2) = {\sqrt 2}iD_{\alpha}^1. \\
\end{aligned}
\tag{4.16}
\end{equation*}
Thus $sp(\psi_i) \cong \hbox{Span} (E_{\alpha}^i, H_{\alpha}^i, F_{\alpha}^i)$
for $i = 1, 2, 3$.
$$\eqno\Q$$

{\it Remark 4.2:}
We will use the following commutation relations:
\begin{equation*}
\begin{aligned}
&[T_{\alpha}^1, T_{\alpha}^3] = E_{\alpha}^1,\quad
[T_{\alpha}^1, D_{\alpha}^4] = -(1+\alpha)F_{\alpha}^2,\\
&[T_{\alpha}^2, T_{\alpha}^4] = E_{\alpha}^1,\quad
[T_{\alpha}^2, D_{\alpha}^3] = (1+\alpha)F_{\alpha}^2,\\
&[D_{\alpha}^1, T_{\alpha}^4] = (1+\alpha)E_{\alpha}^2,\quad
[D_{\alpha}^1, D_{\alpha}^3] = F_{\alpha}^1,\\
&[D_{\alpha}^2, T_{\alpha}^3] = -(1+\alpha)E_{\alpha}^2,\quad
[D_{\alpha}^2, D_{\alpha}^4] = F_{\alpha}^1.
\end{aligned}
\tag{4.17}
\end{equation*}
\vskip 0.2in
\noindent
{\bf V. Superalgebra $\hat{K}'(4)$}
\vskip 0.2in

By definition,
$$K(2N) = \lbrace D \in W(2N)\mid D\Omega  = f\Omega \hbox{ for some }
f\in \Lambda(1,2 N)\rbrace, \eqno (5.1)$$
where
$\Omega = dt + \sum_{i=1}^N \xi_id\eta_i + \eta_id\xi_i$
is a differential 1-form, which is called a {\it contact form},
see Refs. 7-10, 16-18 and 28.
The Euler operator is defined by
$E = \sum_{i=1}^N \xi_i\partial_{\xi_i}
+ \eta_i\partial_{\eta_i}$.
We also define operators $\Delta = 2 - E$ and
$H_f = (-1)^{p(f) +1}\sum_{i=1}^N
\partial_{\xi_i}f
\partial_{\eta_i} +
\partial_{\eta_i}f
\partial_{\xi_i}$, where $f\in \Lambda(1, 2N)$.

There is a one-to-one correspondence between the differential operators
$D\in K(2N)$ and the functions $f \in \Lambda(1, 2N)$.
The correspondence $f \leftrightarrow D_f$ is given by
$$D_f = \Delta(f){\partial \over {\partial t}} +
{\partial f \over {\partial t}}E - H_f.\eqno (5.2)$$
The contact bracket on $\Lambda(1, 2N)$ is
$$\lbrace f, g\rbrace_K = \Delta(f){\partial_t g} -
{\partial_t f}\Delta(g) - \lbrace f, g\rbrace_{P.b},\eqno (5.3)$$
where
$$\lbrace f, g\rbrace_{P.b} =
(-1)^{p(f) +1}\sum_{i=1}^N (\partial_{\xi_i} f
\partial_{\eta_i} g +
\partial_{\eta_i} f \partial_{\xi_i} g)\eqno (5.4)$$
 is the Poisson bracket.
Thus $[D_f, D_g] = D_{{\lbrace f, g \rbrace}_K}$.

The superalgebras $K(2N)$ are simple, except when $N = 2$. If $N = 2$,
then the derived superalgebra $K'(4) = [K(4), K(4)]$ is a simple ideal in
$K(4)$ of codimension one defined from the exact sequence
$$0\rightarrow K'(4)\rightarrow
 K(4)\rightarrow \C
D_{t^{-1} \xi_1\xi_2\eta_1\eta_2}\rightarrow 0. \eqno (5.5)$$
The superalgebra $K'(4)$ has 3 independent central extensions,
see Refs. 7, 10, 29 and 30.
The following statement is proven in  Refs. 5 and 13.

{\it Proposition 5.1:} There exists an embedding
 $$i_0: K'(4)\longrightarrow P(4).\eqno (5.6)$$
The superalgebra $i_0(K'(4))$ is spanned
by the 12 fields:
\begin{equation*}
\begin{aligned}
&L_n = t^{n+1}\tau, \quad Q_n = t^{n+1}\tau\xi_1\xi_2,\\
&X_n^i = t^{n+1}\tau\xi_i, \quad
Y_n^i = t^{n}\eta_i,\\
&R_n^{ji} = t^{n}\xi_j\eta_i, \quad
Z_n^i = t^{n}\xi_1\xi_2\eta_i,
\end{aligned}
\tag{5.7}
\end{equation*}
where $i, j = 1, 2$,
and 4 fields:
\begin{equation*}
\begin{aligned}
&G_n^0 = t^{n-1}\tau^{-1}\eta_1\eta_2,\\
&G_n^i = t^{n-1}\tau^{-1}\xi_i\eta_1\eta_2, \quad i = 1, 2,\\
&G_n^3 = nt^{n-1}\tau^{-1}\xi_1\xi_2\eta_1\eta_2, \quad n\not= 0.
\end{aligned}
\tag{5.8}
\end{equation*}
Note that $L_n$ is a Virasoro field.
Let $\hat{K}'(4) = {K}'(4)\oplus\C C$ be one of three independent central extensions of $K'(4)$, such that the
corresponding 2-cocycle  is
\begin{equation*}
\begin{aligned}
&c(L_n, G_k^3) = -n\delta_{n+k,0},\\
&c(X_n^i, G_k^j) = (-1)^j\delta_{n+k,0}, \hbox{ }1\leq i\not= j\leq 2,\\
&c(Q_n, G_k^0) = \delta_{n+k,0}.
\end{aligned}
\tag{5.9}
\end{equation*}
For each $h\in (0, 1]$, there exists an embedding
$$i_h:\hat{K}'(4)\longrightarrow P_h(4).\eqno (5.10)$$
The superalgebra $i_h(\hat{K}'(4))$ is spanned by the 12 fields (5.7) and
4 fields:
\begin{equation*}
\begin{aligned}
&G_{n, h}^0 = \tau^{-1}\circ_ht^{n-1}\eta_1\eta_2,\\
&G_{n, h}^i = \tau^{-1}\circ_ht^{n-1}\eta_1\eta_2\xi_i, \quad i = 1, 2,\\
&G_{n, h}^3 = n\tau^{-1}\circ_ht^{n-1}\eta_1\eta_2\xi_1\xi_2 + ht^n.
\end{aligned}
\tag{5.11}
\end{equation*}
Note that the central element in $i_h(\hat{K}'(4))$ is $G^3_{0,h} = h$, and
$$\hbox{lim}_{h\rightarrow 0}i_h(\hat{K}'(4)) = i_0(K'(4))\subset P(4). \eqno (5.12)$$

{\bf Theorem 5.2:} Let $\Gamma_{\alpha, h}$, where
$\alpha \in \C$ and $h\in (0, 1]$, be spanned by the following elements in
$P_h(4)$:
\begin{equation*}
\begin{aligned}
&E_{\alpha,h}^1 =  t^2, \quad H_{\alpha,h}^1 = t\tau + {{\alpha + 1}\over 2}h,\\
&F_{\alpha,h}^1 = \tau^2 - \alpha(2t^{-2}\xi_1\xi_2\eta_1\eta_2 +
t^{-2}(\xi_1\eta_1 + \xi_2\eta_2)h - t^{-1}\tau h),\\
&E_{\alpha,h}^2 = \xi_1\xi_2,\quad
F_{\alpha,h}^2 = \eta_1\eta_2,\quad
H_{\alpha,h}^2 =  \xi_1\eta_1 + \xi_2\eta_2 - h,\\
&E_{\alpha,h}^3 = \xi_1\eta_2,\quad
F_{\alpha,h}^3 = \xi_2\eta_1,\quad
H_{\alpha,h}^3 =  \xi_1\eta_1 - \xi_2\eta_2,\\
&T_{\alpha,h}^1 =   t\eta_1,\quad
T_{\alpha,h}^2 =  t\eta_2,\quad
T_{\alpha,h}^3 =  t\xi_1,\quad
T_{\alpha,h}^4 =  t\xi_2,\\
&D_{\alpha,h}^1 =  \tau\xi_1 + \alpha t^{-1}\xi_1\xi_2\eta_2,\quad
D_{\alpha,h}^2 = \tau\xi_2 - \alpha t^{-1}\xi_1\xi_2\eta_1,\\
&D_{\alpha,h}^3 = \tau\eta_1 + \alpha t^{-1}\eta_1\eta_2\xi_2,\quad
D_{\alpha,h}^4 = \tau\eta_2 - \alpha t^{-1}\eta_1\eta_2\xi_1.
\end{aligned}
\tag{5.13}
\end{equation*}
Then $\Gamma_{\alpha,h}\cong
\Gamma(2, -1 -\alpha, \alpha - 1)$,
and $\hbox{lim}_{h\rightarrow 0}\Gamma_{\alpha,h} = \Gamma_{\alpha}\subset P(4)$.

\noindent
{\it Proof.}
We can obtain the second embedding
 $$j_h:\hat{K}'(4)\longrightarrow P_h(4)\eqno (5.14)$$
 for each $h\in (0, 1]$,
 if we interchange $\xi_i$ with $\eta_i$ in all the formulas for the embedding (5.10).
Then
$$\hbox{lim}_{h\rightarrow 0}j_h(\hat{K}'(4)) = j_0(K'(4))\subset P(4). \eqno (5.15)$$
In (4.8) and (4.11)
we obtained embeddings
$$S_{\alpha}^1\subset i_0(K'(4)), \quad   S_{\alpha}^2\subset j_0(K'(4)).$$
Naturally
$$S_{\alpha}^1 = S_{\alpha,h}^1\subset i_h(\hat{K}'(4)).\eqno(5.16)$$
To obtain embedding
$$S_{\alpha,h}^2 \subset j_h(\hat{K}'(4))\eqno (5.17)$$
we interchange $\xi_i$ with $\eta_i$ in all formulas for the embedding (5.16).
Thus
$$S_{\alpha,h}^2 = \hbox{Span}(L_{n, \alpha,h}^2, E_{n, \alpha,h}^3, F_{n, \alpha,h}^3,
H_{n, \alpha,h}^3,
T_{n, \alpha,h}^3, T_{n, \alpha,h}^4, D_{n, \alpha,h}^3,  D_{n, \alpha,h}^4),\eqno (5.18)$$
where
\begin{equation*}
\begin{aligned}
&L_{n, \alpha, h}^2 =
t^{n+1}\tau + {1\over 2}(\alpha + n + 1)(\eta_1\xi_1 + \eta_2\xi_2),\\
&E_{n, \alpha,h}^3 = E_{n, \alpha}^3, F_{n, \alpha,h}^3 = F_{n, \alpha}^3,
H_{n, \alpha,h}^3 = H_{n, \alpha}^3,\\
&T_{n, \alpha,h}^3 = T_{n, \alpha}^3, \quad T_{n, \alpha,h}^4 =  T_{n, \alpha}^4,\\
&D_{n, \alpha, h}^3 = t^n\tau\eta_1 + (\alpha + n) t^{n-1}\eta_1\eta_2\xi_2,\\
&D_{n, \alpha, h}^4 = t^n\tau\eta_2 - (\alpha + n) t^{n-1}\eta_1\eta_2\xi_1.
\end{aligned}
\tag{5.19}
\end{equation*}
For each $\alpha \in \C$ and each $h\in (0, 1]$, the zero modes of the fermionic fields
$$T_{n, \alpha,h}^i \hbox{ and }  D_{n, \alpha,h}^i \hbox{ for } i = 1, 2, 3, 4\eqno (5.20)$$
generate $\Gamma_{\alpha,h}$ and it  is isomorphic to
$\Gamma (2, -1 - \alpha, \alpha -1)$.
One can use the commutation relations  given in  Remark 4.2 to find
$E_{\alpha,h}^i, F_{\alpha,h}^i$ and $H_{\alpha,h}^i$ for $i = 1, 2$.
$$\eqno \Q$$

\vskip 0.2in
\noindent
{\bf VI. Realizations as matrices over a Weyl algebra}
\vskip 0.2in

In this section we will describe
$\hat{K}'(4)$ in terms of matrices over a Weyl algebra.

\noindent
By definition, a Weyl algebra is
$$\W = \sum_{i\geq 0}\A d^i, \eqno (6.1)$$
 where $\A$ is an associative commutative algebra and
 $d:\A\rightarrow \A$ is a derivation of $\A$,
 with the relations
$$da = d(a) + ad, \quad a\in\A,\eqno (6.2)$$
see Refs. 19 and 20. Set
$$\A = \C[t, t^{-1}], \hbox{ } d= L_0 = t\tau. \eqno (6.3)$$
Let $M({2|2}, \W)$ be the Lie superalgebra of $4\times 4$
matrices over  $\W$.

{\bf Theorem 6.1:} There exists an embedding
$$I:\hat{K}'(4)\longrightarrow M({2|2}, \W).\eqno (6.4)$$
The superalgebra $I(\hat{K}'(4))$ is spanned by the following elements:
\bigskip
\begin{equation*}
\begin{aligned}
&I(L_n) = \left(\begin{array}{cc|cc}
dt^n&0&0&0\\
0&t^nd&0&0\\
\hline
0&0&t^nd&0\\
0&0&0&t^nd
\end{array}\right),
 \quad
I(G_n^3) = t^n1_{2|2},\\
&I(R_n^{11}) = \left(\begin{array}{cc|cc}
0&0&0&0\\
0&t^n&0&0\\
\hline
0&0&t^n&0\\
0&0&0&0
\end{array}\right), \quad
I(R_n^{22}) = \left(\begin{array}{cc|cc}
0&0&0&0\\
0&t^n&0&0\\
\hline
0&0&0&0\\
0&0&0&t^n
\end{array}\right),\\
&I(R_n^{12}) = \left(\begin{array}{cc|cc}
0&0&0&0\\
0&0&0&0\\
\hline
0&0&0&t^n\\
0&0&0&0
\end{array}\right), \quad
I(R_n^{21}) = \left(\begin{array}{cc|cc}
0&0&0&0\\
0&0&0&0\\
\hline
0&0&0&0\\
0&0&t^n&0
\end{array}\right),\\
&I(G_n^{0}) = \left(\begin{array}{cc|cc}
0&-t^n&0&0\\
0&0&0&0\\
\hline
0&0&0&0\\
0&0&0&0
\end{array}\right), \quad
I(Q_n) = \left(\begin{array}{cc|cc}
0&0&0&0\\
t^n&0&0&0\\
\hline
0&0&0&0\\
0&0&0&0
\end{array}\right),
\end{aligned}
\tag{6.5}
\end{equation*}
\begin{equation*}
\begin{aligned}
&I(Y_n^{1}) = \left(\begin{array}{cc|cc}
0&0&dt^n&0\\
0&0&0&0\\
\hline
0&0&0&0\\
0&t^n&0&0
\end{array}\right), \quad
I(Y_n^2) = \left(\begin{array}{cc|cc}
0&0&0&dt^n\\
0&0&0&0\\
\hline
0&-t^n&0&0\\
0&0&0&0
\end{array}\right),\\
&I(X_n^{1}) = \left(\begin{array}{cc|cc}
0&0&0&0\\
0&0&0&t^nd\\
\hline
t^n&0&0&0\\
0&0&0&0
\end{array}\right), \quad
I(X_n^2) = \left(\begin{array}{cc|cc}
0&0&0&0\\
0&0&-t^nd&0\\
\hline
0&0&0&0\\
t^n&0&0&0
\end{array}\right),
\end{aligned}
\end{equation*}
\begin{equation*}
\begin{aligned}
&I(G_n^{1}) = \left(\begin{array}{cc|cc}
0&0&0&-t^n\\
0&0&0&0\\
\hline
0&0&0&0\\
0&0&0&0
\end{array}\right), \quad
I(G_n^2) = \left(\begin{array}{cc|cc}
0&0&t^n&0\\
0&0&0&0\\
\hline
0&0&0&0\\
0&0&0&0
\end{array}\right),\\
&I(Z_n^{1}) = \left(\begin{array}{cc|cc}
0&0&0&0\\
0&0&t^n&0\\
\hline
0&0&0&0\\
0&0&0&0
\end{array}\right), \quad
I(Z_n^2) = \left(\begin{array}{cc|cc}
0&0&0&0\\
0&0&0&t^n\\
\hline
0&0&0&0\\
0&0&0&0
\end{array}\right).
\end{aligned}
\end{equation*}
Note that the central element is $C = I(G_0^3) = 1_{2|2}$.

\noindent
{\it Proof.}
Consider the embedding
$$i_h: \hat{K}'(4)\longrightarrow P_h(4).\eqno (6.6)$$
Let $V^{\mu} = t^{\mu}\C[t, t^{-1}]\otimes\Lambda(\xi_1, \xi_2)$, where $\mu\in\C\backslash\Z$.
We fix $h = 1$, and define
a representation of $\hat{K}'(4)$ in $V^{\mu}$ according to the formulas (5.7) and (5.11). Namely,
$\xi_i$ is the operator of multiplication in $\Lambda(\xi_1, \xi_2)$,
$\eta_i$ is identified with $\partial_{\xi_i}$,  $\tau^{-1}$ is identified with an antiderivative,
and the central element $C = 1\in P_{h = 1}(4)$ acts by the identity operator.
Consider the following basis in $V^{\mu}$:
\begin{equation*}
\begin{aligned}
&v_m^0(\mu) = {1\over{m+\mu}}t^{m+\mu},\quad v_m^1(\mu) = t^{m+\mu}\xi_1,\\
&v_m^2(\mu) = t^{m+\mu}\xi_2, \quad v_m^{12}(\mu) = t^{m+\mu}\xi_1\xi_2
\hbox{ for all }m\in\Z.
\end{aligned}
\tag{6.7}
\end{equation*}
Explicitly, the action of $\hat{K}'(4)$ on $V^{\mu}$ is given as follows:
\begin{equation*}
\begin{aligned}
&L_n(v_m^0(\mu)) = (n + m + \mu) v_{m+n}^0(\mu),\\
&L_n(v_m^i(\mu)) = ( m + \mu) v_{m+n}^i(\mu),\quad i = 1, 2, 3,\\
&X_n^i(v_m^0(\mu)) = v_{m+n}^i(\mu),\quad i = 1, 2, \\
&X_n^1(v_m^2(\mu)) = (m + \mu)v_{m+n}^3(\mu), \\
&X_n^2(v_m^1(\mu)) = -(m + \mu) v_{m+n}^3(\mu), \\
&Q_n(v_m^0(\mu)) = v_{m+n}^3(\mu),
\end{aligned}
\tag{6.8}
\end{equation*}
\begin{equation*}
\begin{aligned}
&Y_n^i(v_m^i(\mu)) = (n + m + \mu)v_{m+n}^0(\mu),\quad i = 1, 2,\\
&Y_n^1(v_m^3(\mu)) = v_{m+n}^2(\mu),\\
&Y_n^2(v_m^3(\mu)) =-v_{m+n}^1(\mu),\\
&R_n^{ii}(v_m^i(\mu)) = v_{m+n}^i(\mu),\quad i = 1, 2,\\
&R_n^{ii}(v_m^3(\mu)) = v_{m+n}^3(\mu),\quad i = 1, 2,\\
&R_n^{ij}(v_m^j(\mu)) = v_{m+n}^i(\mu),\quad i \not= j = 1, 2,\\
&Z_n^i(v_m^i(\mu)) = v_{m+n}^3(\mu),\quad i = 1, 2,\\
&G_{n,1}^0(v_m^3(\mu)) = -v_{m+n}^0(\mu),\\
&G_{n,1}^1(v_m^2(\mu)) = -v_{m+n}^0(\mu),\quad
G_{n,1}^2(v_m^1(\mu)) = v_{m+n}^0(\mu),\\
&G_{n,1}^3(v_m^i(\mu)) = v_{m+n}^i(\mu), \quad n\not= 0, \hbox{ } i = 0, 1, 2, 3.
\end{aligned}
\end{equation*}
These formulas remain valid for  $\mu = 0$. Thus we obtain a representation of
$\hat{K}'(4)$ in the superspace $V = \C[t, t^{-1}]\otimes \Lambda(\xi_1, \xi_2)$ with a basis
$$\lbrace v_{m}^0,  v_{m}^3; v_{m}^1,  v_{m}^2 \rbrace,$$ where
$$v_m^0 = t^m,  \quad v_m^3 = t^m\xi_1\xi_2,\quad
v_m^i = t^m\xi_i, \quad i = 1, 2, \quad m\in\Z.$$
We have
\begin{equation*}
\begin{aligned}
&L_n(v_m^0) = dt^{n}v_{m}^0, \quad
L_n(v_m^i) = t^{n}dv_{m}^i, \quad i = 1, 2, 3,\\
&X_n^i(v_m^0) =  t^{n}v_{m}^i,\hbox{ } i = 1, 2, \quad
X_n^1(v_m^2) = t^{n}dv_{m}^3, \\
&X_n^2(v_m^1) = - t^{n}dv_{m}^3, \quad
Q_n(v_m^0) = t^nv_{m}^3, \\
&Y_n^i(v_m^i) = dt^nv_{m}^0,\quad i = 1, 2,\quad
Y_n^1(v_m^3) = t^nv_{m}^2,\\
&Y_n^2(v_m^3) =- t^nv_{m}^1,\quad
R_n^{ii}(v_m^i) = t^nv_{m}^i, \quad i = 1, 2,\\
&R_n^{ii}(v_m^3) = t^nv_{m}^3, \quad
R_n^{ij}(v_m^j) = t^nv_{m}^i, \quad i\not= j = 1, 2,\\
&Z_n^i(v_m^i) = t^nv_{m}^3,\quad i = 1, 2,\\
&G_{n,1}^0(v_m^3) = -t^nv_{m}^0,\quad
G_{n,1}^1(v_m^2) = -t^nv_{m}^0,\quad
G_{n,1}^2(v_m^1) = t^nv_{m}^0,\\
&G_{n,1}^3(v_m^i) = t^nv_{m}^i, \quad n\not= 0,\hbox{ }i = 0, 1, 2, 3.
\end{aligned}
\tag{6.9}
\end{equation*}
Thus we obtain the above-mentioned  realization of $\hat{K}'(4)$ as a subsuperalgebra of matrices of size $4\times 4$ over $\W$.
$$\eqno \Q$$

{\it Remark 6.2.}
Naturally,  $V = \oplus_m V_m$, where
$V_m = t^{m}\otimes \Lambda(\xi_1, \xi_2)$.
Recall that the element $L_0 = t\tau$ of the Virasoro algebra defines
a $\Z$-grading in $\hat{K}'(4)$:  $\hat{K}'(4) = \oplus_i\g_i$.
It follows from (6.8) that
$$\g_i(V_m) \subset V_{m+i}. $$
Note that $\g_0$ is isomorphic to the universal central extension
of ${sl}(2|2)$, and it is  realized as a superalgebra of $4\times 4$ matrices
over $\W$ of type
$$ \left( \begin{array}{cc}
A & B + d\tilde{C}\\
C & D
\end{array}\right)\oplus \C d \cdot 1_{2|2},\eqno (6.10)$$
where $A, B, C, D \in gl(2, \C)$ and $trA = tr D$.
$\tilde{C}$ is determined by the following
conditions:
\begin{equation*}
\begin{aligned}
&\hbox{if } C = E_{ii}, \hbox{ then } \tilde{C} = E_{jj}, \hbox{ where } i\not= j,\\
&\hbox{if } C = E_{ij}, \hbox{ } i\not= j,
\hbox{ then } \tilde{C} = -E_{ij},
\end{aligned}
\tag{6.11}
\end{equation*}
where $E_{ij}$ is an elementary $2\times 2$-matrix.

\vskip 0.1in

It  was observed in  Refs. 30 and 31 that
the big $N = 4$ superconformal algebra contains $D(2, 1;\alpha)$
as a subsuperalgebra. In the next theorem, we give a realization of
$D(2, 1;\alpha)$ inside $I(\hat{K}'(4))$.
Note that it is different from the realization of $D(2, 1;\alpha)$
inside $M(2|2, \W)$, which one can directly obtain from Theorem 5.2.

\vskip 0.1in

{\bf Theorem 6.3:} For each
$\alpha \in \C$ the superalgebra
$\Gamma (2, -1 - \alpha, \alpha - 1)$
is realized inside the superalgebra $I(\hat{K}'(4))$
as follows:
\bigskip
\begin{equation*}
\begin{aligned}
&T_{\alpha}^{1} = \left(\begin{array}{cc|cc}
0&0&t(d+1)&0\\
0&0&0&0\\
\hline
0&0&0&0\\
0&t&0&0
\end{array}\right), \quad
T_{\alpha}^2 = \left(\begin{array}{cc|cc}
0&0&0&t(d+1)\\
0&0&0&0\\
\hline
0&-t&0&0\\
0&0&0&0
\end{array}\right),\\
&D_{\alpha}^{1} = \left(\begin{array}{cc|cc}
0&0&0&0\\
0&0&0&t^{-1}(d + \alpha)\\
\hline
t^{-1}&0&0&0\\
0&0&0&0
\end{array}\right), \quad
D_{\alpha}^2 = \left(\begin{array}{cc|cc}
0&0&0&0\\
0&0&-t^{-1}(d +\alpha)&0\\
\hline
0&0&0&0\\
t^{-1}&0&0&0
\end{array}\right),\\
&T_{\alpha}^{3} = \left(\begin{array}{cc|cc}
0&0&0&0\\
0&0&0&t(d+1)\\
\hline
t&0&0&0\\
0&0&0&0
\end{array}\right), \quad
T_{\alpha}^4 = \left(\begin{array}{cc|cc}
0&0&0&0\\
0&0&-t(d+1)&0\\
\hline
0&0&0&0\\
t&0&0&0
\end{array}\right),\\
&D_{\alpha}^{3} = \left(\begin{array}{cc|cc}
0&0&t^{-1}(d+\alpha)&0\\
0&0&0&0\\
\hline
0&0&0&0\\
0&t^{-1}&0&0
\end{array}\right), \quad
D_{\alpha}^4 = \left(\begin{array}{cc|cc}
0&0&0&t^{-1}(d+\alpha)\\
0&0&0&0\\
\hline
0&-t^{-1}&0&0\\
0&0&0&0
\end{array}\right),
\end{aligned}
\tag{6.12}
\end{equation*}
\begin{equation*}
\begin{aligned}
&E_{\alpha}^{1} = \left(\begin{array}{c|c}
t^2(d+2) 1_2&0\\
\hline
0&t^2(d+1) 1_2\\
\end{array}\right), \\
&F_{\alpha}^1 = \left(\begin{array}{c|c}
 t^{-2}(d + \alpha - 1)1_2&0\\
\hline
0&t^{-2}(d +\alpha)1_2\\
\end{array}\right),\\
&H_{\alpha}^{1} =
(d + {{1 +\alpha}\over 2})1_{2|2},
\end{aligned}
\end{equation*}
\begin{equation*}
\begin{aligned}
&E_{\alpha}^{2} = \left(\begin{array}{cc|cc}
0&0&0&0\\
1&0&0&0\\
\hline
0&0&0&0\\
0&0&0&0
\end{array}\right),
F_{\alpha}^2 = \left(\begin{array}{cc|cc}
0&-1&0&0\\
0&0&0&0\\
\hline
0&0&0&0\\
0&0&0&0
\end{array}\right),
H_{\alpha}^{2} = \left(\begin{array}{cc|cc}
-1&0&0&0\\
0&1&0&0\\
\hline
0&0&0&0\\
0&0&0&0
\end{array}\right),\\
&E_{\alpha}^{3} = \left(\begin{array}{cc|cc}
0&0&0&0\\
0&0&0&0\\
\hline
0&0&0&1\\
0&0&0&0
\end{array}\right),
F_{\alpha}^3 = \left(\begin{array}{cc|cc}
0&0&0&0\\
0&0&0&0\\
\hline
0&0&0&0\\
0&0&1&0
\end{array}\right),
H_{\alpha}^{3} = \left(\begin{array}{cc|cc}
0&0&0&0\\
0&0&0&0\\
\hline
0&0&1&0\\
0&0&0&-1
\end{array}\right).
\end{aligned}
\end{equation*}
{\it Proof.}
Consider the  embedding
$$j_h:\hat{K}'(4)\longrightarrow P_h(4).\eqno (6.13)$$
 To describe the associated representations we choose the following
 basis in $V^{\mu}$:
\begin{equation*}
\begin{aligned}
&v_m^0(\mu) = t^{m+\mu},\quad v_m^1(\mu) = t^{m+\mu}\xi_1,\\
&v_m^2(\mu) = t^{m+\mu}\xi_2, \quad v_m^{12}(\mu) = {1\over{m+\mu}}t^{m+\mu}\xi_1\xi_2, \quad m\in\Z.
\end{aligned}
\tag{6.14}
\end{equation*}
One can repeat the construction given in the proof of Theorem 6.1,
and obtain an embedding
$$J: \hat{K}'(4)\longrightarrow M({2|2}, \W).\eqno (6.15)$$
Note that under both embeddings, $\hat{K}'(4)$ is realized
as the same matrix superalgebra:
$$I(\hat{K}'(4)) = J(\hat{K}'(4)).\eqno (6.16)$$
We have
\begin{equation*}
\begin{aligned}
&J(L_n) = I(L_n) - nI(G_n^3) + nI(R_n^{11}) + nI(R_n^{22}), \quad J(Q_n) = I(G_n^0),\\
&J(R_n^{11}) = I(G_n^3) - I(R_n^{11}), \quad
J(R_n^{22}) = I(G_n^3) - I(R_n^{22}),\\
&J(R_n^{12}) =  - I(R_n^{21}), \quad
J(R_n^{21}) = - I(R_n^{12}), \\
&J(G_n^{0}) = I(Q_n), \quad J(G_n^{3}) = I(G_n^3),\\
&J(Y_n^1) = I(X_n^1) + nI(Z_n^2), \quad J(Y_n^2) = I(X_n^2) - nI(Z_n^1),\\
&J(X_n^1) = I(Y_n^1) - nI(G_n^2), \quad J(X_n^2) = I(Y_n^2) + nI(G_n^1),\\
&J(G_n^1) = I(Z_n^1), \quad J(G_n^2) = I(Z_n^2), \quad
J(Q_n^1) = I(G_n^1), \quad J(Q_n^2) = I(G_n^2).
\end{aligned}
\tag{6.17}
\end{equation*}
To find matrix realizations of
$T_{\alpha}^i$ and  $D_{\alpha}^i$, we use (6.4), if $i = 1, 2$,
and we use (6.15), if $i = 3, 4$.
Note that  formulas (6.4) and (6.15) determine
the same matrices for $E_{\alpha}^3, F_{\alpha}^3$ and  $H_{\alpha}^3$.
Finally, to find $E_{\alpha}^i, F_{\alpha}^i$ and $H_{\alpha}^i$ for $i = 1, 2$,
we use the commutation relations given in Remark 4.2, and the relations in $\W$:
$$dt^n = t^nd +  nt^n \hbox{ for all } n\in\Z.\eqno (6.18)$$
Thus we have
\begin{equation*}
\begin{aligned}
&T_{\alpha}^1 = I(Y_1^1), \quad T_{\alpha}^2 = I(Y_1^2),\\
&T_{\alpha}^3 = I(X_1^1) + I(Z_1^2), \quad T_{\alpha}^4 = I(X_1^2) - I(Z_1^1),\\
&D_{\alpha}^1 = I(X_{-1}^1) + \alpha I(Z_{-1}^2),
\quad D_{\alpha}^2 = I(X_{-1}^2) - \alpha I(Z_{-1}^1),\\
&D_{\alpha}^3 = I(Y_{-1}^1) + (\alpha + 1)I(G_{-1}^2),
 \quad D_{\alpha}^4 = I(Y_{-1}^2) - (\alpha + 1)I(G_{-1}^1),\\
&E_{\alpha}^1 = I(L_2) + I(R_{2}^{11}) + I(R_{2}^{22}), \\
&F_{\alpha}^1 = I(L_{-2})  + (\alpha + 1)I(G_{-2}^{3}) - I(R_{-2}^{11})- I(R_{-2}^{22}),\\
&H_{\alpha}^1 = I(L_0) + {1\over 2}(1 + \alpha)C,\\
&E_{\alpha}^2 = I(Q_0),\quad F_{\alpha}^2 = I(G_0^0),
\quad H_{\alpha}^2 = I(R_0^{11}) + I(R_0^{22}) - C,\\
&E_{\alpha}^3 = I(R_0^{12}),\quad F_{\alpha}^3 = I(R_0^{21}),
\quad H_{\alpha}^3 = I(R_0^{11}) - I(R_0^{22}).
\end{aligned}
\tag{6.19}
\end{equation*}
$$\eqno \Q$$

{\it Remark 6.4:}
In Theorem 5.2 we described an embedding of
 $\Gamma (2, -1 - \alpha, \alpha - 1)$ into $P_h(4)$.
 Note that it is actually an embedding of $\Gamma (2, -1 - \alpha, \alpha - 1)$
 into the Lie superalgebra of {\it differential operators} on $S^{1|2}$.
One can use the fields in (5.7) and (5.11) and formulas
(6.19) to obtain a different  embedding of this superalgebra into $P_h(4)$ such that
$$\Gamma_{\alpha, h}\subset i_h(\hat{K}'(4))\eqno (6.20)$$
for each $h\in (0, 1]$ and each $\alpha\in\C$.
In this embedding the pseudodifferential symbols are essentially used.
$\Gamma_{\alpha, h}\cong \Gamma (2, -1 - \alpha, \alpha - 1)$ is spanned by the
following elements:
\begin{equation*}
\begin{aligned}
&E_{\alpha,h}^1 =  t^3\tau + t^2(\xi_1\eta_1 +  \xi_2\eta_2),\\
&F_{\alpha,h}^1 = t^{-1}\tau +
(\alpha + 1)(-2\tau^{-1}\circ_ht^{-3}\eta_1\eta_2\xi_1\xi_2 + ht^{-2})
-t^{-2}(\xi_1\eta_1 +  \xi_2\eta_2),\\
&H_{\alpha,h}^1 = t\tau + {{\alpha + 1}\over 2}h,\\
&E_{\alpha,h}^2 = t\tau\xi_1\xi_2,\quad
F_{\alpha,h}^2 = \tau^{-1}\circ_ht^{-1}\eta_1\eta_2,\quad
H_{\alpha,h}^2 =  \xi_1\eta_1 + \xi_2\eta_2 - h,\\
&E_{\alpha,h}^3 = \xi_1\eta_2,\quad
F_{\alpha,h}^3 = \xi_2\eta_1,\quad
H_{\alpha,h}^3 =  \xi_1\eta_1 - \xi_2\eta_2,
\end{aligned}
\tag{6.21}
\end{equation*}
\begin{equation*}
\begin{aligned}
&T_{\alpha,h}^1 =   t\eta_1,\quad
T_{\alpha,h}^2 =  t\eta_2,\\
&T_{\alpha,h}^3 =  t^2\tau\xi_1 + t\xi_1\xi_2\eta_2,\quad
T_{\alpha,h}^4 =  t^2\tau\xi_2- t\xi_1\xi_2\eta_1,\\
&D_{\alpha,h}^1 =  \tau\xi_1 + \alpha t^{-1}\xi_1\xi_2\eta_2,\quad
D_{\alpha,h}^2 = \tau\xi_2 - \alpha t^{-1}\xi_1\xi_2\eta_1,\\
&D_{\alpha,h}^3 = t^{-1}\eta_1 +
(\alpha + 1)\tau^{-1}\circ_ht^{-2}\eta_1\eta_2\xi_2,\\
&D_{\alpha,h}^4 = t^{-1}\eta_2 - (\alpha + 1)\tau^{-1}\circ_ht^{-2}\eta_1\eta_2\xi_1.
\end{aligned}
\end{equation*}
Then
$$\hbox{lim}_{h\rightarrow 0} \Gamma_{\alpha, h} = \Gamma_{\alpha}
\subset i_0(K'(4))\subset P(4). \eqno (6.22)$$
$\Gamma_{\alpha}\cong \Gamma (2, -1 - \alpha, \alpha - 1)$ is spanned by the following elements:
\begin{equation*}
\begin{aligned}
&E_{\alpha}^i = E_{\alpha,h}^i, \quad i = 1, 2, 3, \quad
F_{\alpha}^3 = F_{\alpha,h}^3, \quad
H_{\alpha}^3 = H_{\alpha,h}^3,\\
&T_{\alpha}^i = T_{\alpha,h}^i, \quad i = 1, 2, 3, 4,\quad
D_{\alpha}^i = D_{\alpha,h}^i, \quad i = 1, 2,\\
&F_{\alpha}^1 = t^{-1}\tau
-2(\alpha + 1)t^{-3}\tau^{-1}\xi_1\xi_2\eta_1\eta_2
-t^{-2}(\xi_1\eta_1 +  \xi_2\eta_2),\quad
H_{\alpha}^1 = t\tau, \\
&F_{\alpha}^2 = t^{-1}\tau^{-1}\eta_1\eta_2,\quad
H_{\alpha}^2 =  \xi_1\eta_1 + \xi_2\eta_2,\\
&D_{\alpha}^3 = t^{-1}\eta_1 +
(\alpha + 1)t^{-2}\tau^{-1}\eta_1\eta_2\xi_2,\quad
D_{\alpha}^4 = t^{-1}\eta_2 - (\alpha +1)t^{-2}t^{-2}\tau^{-1}\eta_1\eta_2\xi_1.
\end{aligned}
\tag{6.23}
\end{equation*}
Note that the matrix realization of $\Gamma (2, -1 - \alpha, \alpha - 1)$
in Theorem 6.3 is associated to (6.20), where $h = 1$:
it is
the restriction of the mapping $I$ given in (6.4),
to $\Gamma (2, -1 - \alpha, \alpha - 1)$.

\vskip 0.1in

{\it Remark 6.5:} Recall that superalgebras $\Gamma (2, -1 - \alpha, \alpha - 1)$
are not simple when $\alpha = 1$ or $-1$. Correspondingly,
we have the following realizations of $psl(2|2) = sl(2|2)/<1_{2|2}>$ as
a subsuperalgebra of $M(2|2, \W)$.
If $\alpha = 1$, then
\begin{equation*}
\begin{aligned}
&\hbox{Span} (E^i_{\alpha}, H^i_{\alpha}, F^i_{\alpha},  T^j_{\alpha}, D^j_{\alpha}
\hbox{ }|\hbox{ } i = 1, 2 \hbox{ and } j = 1, \ldots, 4) \cong psl(2|2),\\
&\Gamma (2, -2, 0)/psl(2|2) \cong sl(2).
\end{aligned}
\tag{6.24}
\end{equation*}
If $\alpha = - 1$, then
\begin{equation*}
\begin{aligned}
&\hbox{Span} (E^i_{\alpha}, H^i_{\alpha}, F^i_{\alpha},  T^j_{\alpha}, D^j_{\alpha}
\hbox{ } |\hbox{ } i = 1, 3 \hbox{ and } j = 1, \ldots, 4) \cong psl(2|2),\\
&\Gamma (2, 0, -2)/psl(2|2) \cong sl(2).
\end{aligned}
\tag{6.25}
\end{equation*}

\vskip 0.2 in
\noindent
{\bf Acknowledgments}
\vskip 0.2in

This material is based upon work supported by the National Science Foundation under agreement
$No.\hbox{ } DMS-0111298$. Any opinions, findings and conclusions or recommendations expressed in this material are those of the author and do not necessarily reflect the views of the National Science Foundation.

The author is grateful to the
Institute for Advanced Study for the hospitality and support during term II of the academic year 2006--2007.

She wishes to thank M. G{\"u}naydin for very useful references.

\begin {itemize}

\font\red=cmbsy10
\def\~{\hbox{\red\char'0016}}

\item[{[1]}]
V. G. Kac,
Adv. Math. {\bf 26}, 8 (1977).


\item[{[2]}] M. Scheunert, W. Nahm, and V. Rittenberg,
J. Math. Phys. {\bf 17}, 1640 (1976).


\item[{[3]}] M. Scheunert,
Lecture Notes in Mathematics {\bf 716},  Springer, Berlin (1979).


\item[{[4]}] M. G{\"u}naydin,
Modern Physics Letters A, {\bf 6},  3239 (1991).


\item[{[5]}] E. Poletaeva,
J. Math. Phys. {\bf 42}, 526 (2001); hep-th/0011100
and references therein.


\item[{[6]}] A. Schwimmer and N. Seiberg,
Phys. Lett. B {\bf 184}, 191 (1987).

\item[{[7]}]
V. G. Kac and J. W. van de Leur,
in {\it Strings-88}, edited by S. J. Gates {\it et al}.
(World Scientific, Singapore, 1989), pp 77-106.


\item[{[8]}]
V. G. Kac,
Comm. Math. Phys. {\bf 186}, 233 (1997); {\bf 217}, 697 (2001).


\item[{[9]}]
S.-J. Cheng and V. G.  Kac,
Comm. Math. Phys. {\bf 186},  219 (1997).


\item[{[10]}] P. Grozman, D. Leites, and I. Shchepochkina,
Acta Math. Vietnam. {\bf 26}, 27 (2001); hep-th/9702120.


\item[{[11]}] M. Ademollo, L. Brink, A. D'Adda {\it et al}.,
Phys. Lett. B {\bf 62}, 105 (1976).

\item[{[12]}] M. Ademollo, L. Brink, A. D'Adda {\it et al}.,
Nucl. Phys. B {\bf 111}, 77 (1976).



\item[{[13]}] E. Poletaeva,
J. Math. Phys. {\bf 46}, 103504 (2005).
Publisher's note, J. Math. Phys. {\bf 47}, 019901  (2006);
hep-th/0311247.


\item[{[14]}] E. Poletaeva,
Dynamics of Continuous, Discrete and Impulsive Systems-Series A
(Special Issue), to appear in 2007; arXiv:0707.3097.



\item[{[15]}]
V. G. Kac,
Adv. Math. {\bf 139}, 1 (1998).


\item[{[16]}]
V. G. Kac,
Transform. Groups {\bf 4}, 219 (1999).


\item[{[17]}] I. Shchepochkina,
Funkt. Anal. Priloz. {\bf 33},  59 (1999).
[Funct. Anal. Appl. {\bf 33},  208 (1999)].


\item[{[18]}] I. Shchepochkina,
Represent. Theory {\bf 3}, 373 (1999).


\item[{[19]}] C. Martinez and E. I. Zelmanov,
J. Algebra {\bf 236}, 575 (2001).


\item[{[20]}] C. Martinez and E. I. Zelmanov,
Proc. Natl. Acad. Sci. U.S.A {\bf 100},  8130 (2003).


\item[{[21]}] P. Bowcock, B. L. Feigin, A. M. Semikhatov, and A. Taormina,
 Commun. Math. Phys. {\bf 214},  495 (2000).


\item[{[22]}]
V. G. Kac,
{\it Vertex Algebras for Beginners},
University Lecture Series, Vol. 10 (AMS, Providence, RI, 1996)
(second edition, 1998).


\item[{[23]}]
V. G. Kac,
{\em Proceedings of the International Congress of Mathematicians, Vol. I, Beijing, 2002} (Higher Ed. Press, Beijing, 2002), pp 319-344.


\item[{[24]}]
O. S. Kravchenko and B. A. Khesin,
Funct. Anal. Appl. {\bf 25},  83 (1991).


\item[{[25]}]
B. Khesin, V. Lyubashenko, and C. Roger,
J. Funct. Anal. {\bf 143},  55 (1997).


\item[{[26]}]
V. Ovsienko and C. Roger,
Comm. Math. Phys. {\bf 198}, 97 (1998).


\item[{[27]}]
V. Ovsienko and C. Roger,
Am. Math. Soc. Transl. {\bf 194}, 211 (1999).


\item[{[28]}] B. Feigin and D. Leites,
in {\it Group-Theoretical Methods in
Physics}, edited by  M. Markov {\it et al.},
(Nauka, Moscow, 1983), Vol. 1, 269-273.
[English translation  Gordon and Breach, New York, 1984].


\item[{[29]}] K. Schoutens,
Phys. Lett. B {\bf 194}, 75 (1987).


\item[{[30]}] K. Schoutens,
Nucl. Phys. B {\bf 295}, 634 (1988).


\item[{[31]}] A. Sevrin, W. Troost and A. Van Proeyen,
Phys. Lett. B {\bf 208}, 447 (1988).

\end{itemize}

\end{document}